\documentclass[onecolumn,journal,draftclsnofoot]{IEEEtran}
\usepackage{amsmath,amsfonts}
\usepackage{algorithmic}
\usepackage{algorithm}
\usepackage{array}
\usepackage[caption=false,font=normalsize,labelfont=sf,textfont=sf]{subfig}
\usepackage{textcomp}
\usepackage{stfloats}
\usepackage{url}
\usepackage{verbatim}
\usepackage{graphicx}
\usepackage{cite}
\usepackage{xcolor} 
\usepackage{dirtytalk}

\newtheorem{remark}{Remark}

\begin{document}

\title{An Exploration-Exploitation Approach to Anti-lock Brake Systems}

\author{Benjamin Sullivan, Jingjing Jiang,~\IEEEmembership{Member,~IEEE,}  Georgios Mavros, Wen-Hua Chen,~\IEEEmembership{Fellow,~IEEE,}
\thanks{
This work was supported by the UK Engineering and Physical Sciences Research Council (EPSRC) Established Career Fellowship ‘‘Goal-Oriented Control Systems: Disturbance, Uncertainty and Constarints’’ under the grant number EP/T005734/1.

The authors are with the Department of Aeronautical and Automotive Engineering, Loughborough University, LE11 3TU, UK. Correspondence; b.sullivan@lboro.ac.uk; j.jiang2@lboro.ac.uk;  g.mavros@lboro.ac.uk; w.chen@lboro.ac.uk.

This work has been submitted to the IEEE for possible publication. Copyright may be transferred without notice, after which this version may no longer be accessible.
}}



\maketitle

\begin{abstract}
Anti-lock Brake System (ABS) is a mandatory active safety feature on road vehicles with analogous systems for aircraft and locomotives. This feature aims to prevent locking of the wheels when braking and to improve the handling performance, as well as reduce stopping distance of the vehicle. Estimation uncertainties in the vehicle state and environment (road surface) are often neglected or handled separately from the ABS controller, leading to sub-optimal braking. 
In this paper, a Dual Control for Exploration-Exploitation (DCEE) approach is taken toward the ABS problem which achieves both accurate state (and environment) estimation and superior braking performance. 
Compared with popular Extremum Seeking methods, improvements of up to $15\%$ and $8.5\%$ are shown in stopping time and stopping distance, respectively.  
A Regularized Particle Filter with Markov Chain Monte Carlo step is used to estimate vehicle states and parameters of the Magic Formula tyre model that includes the peak friction coefficient for the environment. 
The effectiveness of the DCEE approach is demonstrated across a range of driving scenarios such as low and high speeds; snow, wet and dry roads and changing road surfaces.
\end{abstract}


\section{Introduction}
\begin{figure}
    \centering
    \includegraphics[scale=0.75]{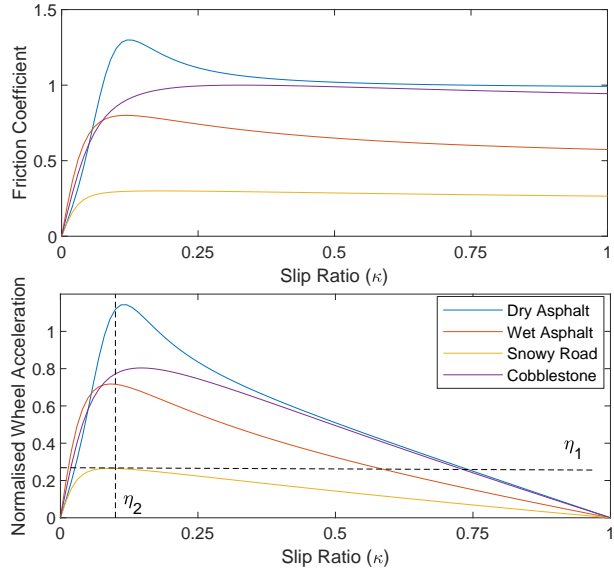}
    \caption{The relationships between wheel slip and friction coefficient (top), as well as between wheel slip and  normalised wheel acceleration (bottom).}
    \label{fig:slipmu_norm}
\end{figure}

\IEEEPARstart{A}{ctive} safety systems, such as Electronic Stability Control, Traction Control and Adaptive Cruise Control, are safety features found on many ground vehicles due to the ability to actively improve controllability and safety of the vehicle. This is either through autonomously activating the vehicle controls (Advanced Driver Assistance Systems in ground vehicles),  or automatically activating inline with direct driver operation, \emph{e.g.} pressing the accelerator may also activate Traction Control. This paper focuses on arguably the most critical active safety feature, the Anti-lock Brake System (ABS). 
The performance of ABS becomes limited when either the state of the vehicle (\emph{i.e.} tyre slip) is poorly estimated or the quality of the road surface is unknown: too much braking causes the wheels to lock or too little braking leads to an increased chance of collision. The relationship between friction coefficient and the vehicle state (wheel slip ratio) is shown in Fig. \ref{fig:slipmu_norm} \cite{Xu2020}\cite{Savaresi2010a}.
The aim of our design is to  achieve the greatest braking performance, \emph{i.e.}, the maximum friction force from the road surface. Thus the ABS problem is twofold: estimate the vehicle states and parameters, as well as control the vehicle to achieve its maximum retardation.

Traditionally, two control methodologies exist for ABS.
They are Wheel Slip Control (WSC) and Wheel Deceleration Control (WDC). 
In the latter class of methods, wheel deceleration control is achieved using rule based heuristics often employing hybrid automata to regulate the wheel deceleration \cite{Gerard2012,Pasillas2006,Kuo1992}. 
This methodology is sub-optimal and in fact attempt to bound the vehicle within the optimal acceleration setpoint using a set of acceleration thresholds. For example, as shown in Fig.~\ref{fig:slipmu_norm}, a hypothetical set-point of wheel acceleration \(\eta_1\) leads to an operating point that is only suitable for snow covered roads, all the other road types lead to the vehicle either sliding significantly or under-braking. 
\IEEEpubidadjcol
WSC methods however, regulate the wheel slip to its largest value before the vehicle begins sliding, achieving the greatest retardation forces. Such methods are beneficial since they have a clear mathematical background and work even if there is no well defined maxima \cite{Pasillas-Lepine2006}.
\cite{Mirzaeinejad2010,Mi2005,Harifi2008} proposed control designs using a priori knowledge of the optimum slip ratio and neglecting measurement noise, while \cite{Rattasiri2007,Zhao2006, Mirzaeinejad2014} estimates the slip value of the vehicle for changing road surfaces. 
Sliding Mode Control has traditionally been used for ABS controllers \cite{Tan1990,Chin1992,Haskara2000} and extended to uncertain systems in \cite{Drakunov1995} where the optimum slip ratio has been estimated in an online fashion using Extremum Seeking \cite{KarticB.Ariyur2004,Dincmen2010,Dincmen2014}. 
Despite this, Sliding Mode Control is prone to high frequency oscillations (chattering) as the controller attempts to hold onto the surface and can cause the vehicle to slide \cite{Mirzaeinejad2014}.

 In this paper, we develop a chatter free solution, namely the Dual Control for Exploration and Exploitation (DCEE) approach which aims to predicatively estimate and achieve the best control action on any road surface.
 This problem is fundamentally an exploration-exploitation of the road surface friction thus the DCEE is qualified.
 The DCEE methodology further improves on the current state of the art by informing the action selection by predicted measurements. 
 Ultimately, this method predicts the future outcome in a cost function that minimizes the error between the predicted tyre force and the maximum available tyre force for any given road surface. 
It is able to deal with time-varying road conditions, which is different from other works  that only consider a time-invariant setpoint \emph{e.g.} \cite{Shaohua2022}.


Kalman Filtering (KF) and subsequent extensions (e.g Extended \cite{Zhang2014,Wenzel2007,Li2021} or Unscented \cite{Bogdanski2017,Chen2016,Qin2022}) have been used to solve the estimation problems (e.g. wheel speed and vehicle velocity), yet do not adequately handle the highly non-linear dynamics of braking systems and non-Gaussian noises.
Limited cases exist for this problem using the Sequential Monte Carlo (SMC) approach \cite{Bogdanski2017}. 
However, such approach cannot deal with higher-dimensional static parameter estimation in this work. 
A GPF approach was implemented to identify a tyre model along with the state of the vehicle \cite{Wang2020}. 
However, this required measurements of lateral and longitudinal velocities, accelerations and wheel speed. 
Hence, the implementation  cost is higher than that of typical WSC or WDC solutions that measure wheel speed and vehicle longitudinal velocity only. 
Furthermore a unique piece-wise tyre force model is identified during this study but lacks usefulness for other systems since it makes assumptions about linearity, saturation and introduces scaling gains. 
Bayesian Inference has shown to perform well in vehicle localization and target tracking for non-linear systems and/or non-Guassian observations for state estimation {\emph{e.g.}\cite{Wenzel2007,Huang2018}} but remains challenging when applied for parameter estimation. We show that a Regularized Particle Filter (RPF) approach for joint state and parameter estimation of unknown vehicle states and tyre model parameters is suitable for the ABS problem. 
The application of the Particle Filter when used for vehicular active safety is under-examined, particularly for dynamic parameters as is the case for changing road surfaces which we demonstrate our method for.


A novel control scheme based on exploitation and exploration has been proposed and verified with simulation, results of which shows an improvement of up to $15\%$ in stopping time and $8.5\%$ in stopping distance by comparing with benchmarks.
We demonstrate that the RPF can estimate both vehicle states and all the parameters of the Magic formula tyre model simultaneously. Furthermore, a novel resampling method is introduced, namely Retrogressive resampling that significantly improves the estimation and control result in the presence of step changes to parameters. These phenomena are typical of braking on changing road surfaces e.g. traversing newly paved road to an old road surface, onto untreated icy roads or over 'potholes'.

\begin{figure}
    \centering
    \includegraphics[scale=0.6]{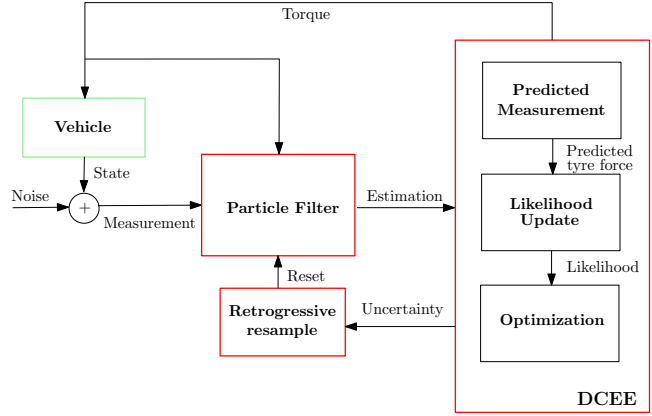}
    \caption{DCEE for ABS system architecture diagram with plant model.}
    \label{fig:dceearch}
\end{figure}

The structure of the rest of the paper is as follows with the system architecture  shown in Fig.~\ref{fig:dceearch}. 
The system model and slip-friction relationship are described in Section \ref{systemdynamics}, while  the joint state and parameter estimation using Particle Filtering is presented in Section \ref{stateparamest} with an effective resampling method for parameter step changes, namely Retrogressive resampling. 
Then a controller is presented using the Dual Control for Exploration and Exploitation method in Section \ref{DCEE}. 
Simulation results are presented in Section \ref{simulation}, demonstrating the effectiveness of the proposed method.
Finally, conclusions and discussion on future progression are conducted in Section \ref{conclusion}.

\section{System Dynamics}\label{systemdynamics}
This work focuses on the longitudinal only behaviour of the vehicle during braking as one considers the lateral vehicle stability to be outside the remit of an ABS feature. Thus we target the 7 degree of freedom (DOF) vehicle model for demonstration of this work.

\subsection{Vehicle Dynamics: 7 DOF Model}
\begin{figure}
    \centering
    \includegraphics[scale=0.68]{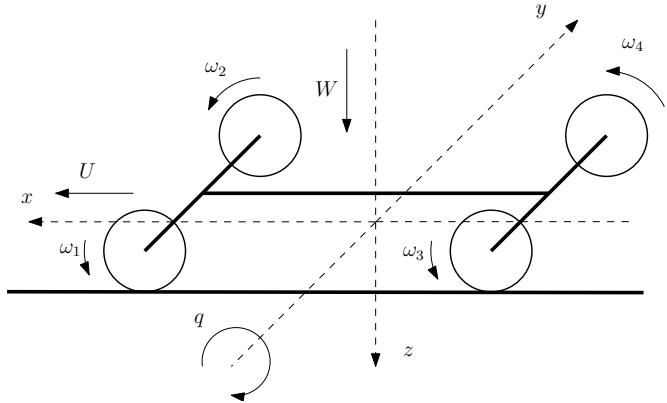}
    \caption{The 7 DOF vehicle dynamics model.}
    \label{fig:7dof}
\end{figure}

The 7 DOF vehicle dynamics model is shown in Fig.~\ref{fig:7dof}. 
From first principles, the summation of all forces  and moments are found using the center of mass  \([x_G,y_G,z_G]\) and align with the SAE frame of reference \cite{Dixon1996}:
\begin{equation}\label{eqn:fx}
\sum^4_{j=1} F_{x,j}=m \left( \frac{dU}{dt} + Wq \right)-m \left(x_G q^2 - z_G \frac{dq}{dt}\right)
\end{equation}
\begin{equation}\label{eqn:fy}
\sum^4_{j=1} F_{z,j} =
m \left( \frac{dW}{dt} - Uq \right) - m \left( z_G q^2 + x_G \frac{dq}{dt}\right)
\end{equation}
\begin{equation}\label{eqn:my}
\sum^4_{j=1} M_y = I_{y y} \frac{dq}{dt} + m z_G \left( \frac{dU}{dt} + Wq \right) - m x_G \left( \frac{dW}{dt} - U q \right)
\end{equation}
where $F_{x,j}$ and $F_{z,j}$ denotes the longitudinal and vertical tyre force at wheel \(j \in \left[ 1,2,3,4 \right]\) as shown in Fig.~\ref{fig:7dof} respectively, while $q$ represents the pitch velocity about the centre of gravity. 
In addition, \(m\) is the total vehicle mass and \(U\) and \(W\) are the longitudinal (forward) and vertical vehicle body centred velocities. 
\(I_{yy}\) is the vehicle body products of inertia about the \(yy\) axis, while $M_y$ denotes the pitching moment.

We model the suspension using a spring/damper, with spring stiffness \(k\) and damping coefficient \(c\), connecting each wheel to the rigid chassis as: 
\begin{equation}\label{eqn:sus1}
F_{z,f}=k_{\mathrm{f}}\left(z-|a|\phi\right)+c_{\mathrm{f}}\left(W-|a| q\right)
\end{equation}
\begin{equation}\label{eqn:sus2}
F_{z,r}=k_{\mathrm{r}}\left(z+|b|\phi\right)+c_{\mathrm{r}}\left(W+|b| q\right)
\end{equation}
where \(\phi\) is the pitch angle, \(a\) and \(b\) are the distances to the centre of gravity from the front and rear axles respectively, \(z\) is the vertical displacement of the centre of gravity.
Note that subscripts \(f\) and \(r\) are used to denote the front and rear axles.

Finally, the remaining four degrees of freedom are from the wheel accelerations, \(\dot{\omega}\), given by the following relationship:
\begin{equation} \label{eqn:wheeldyn}
    I_w\dot{\omega}_j = T_j - RF_{T,j}
\end{equation}
where \(T_j\) denotes the sum of the brake or driving action (control input) applied to the centre of $j$th wheel for all $j\in\{1,2,3,4\}$.
Note that $T_3=T_4=0$ as we only apply braking and driving torques to the front wheels in this study. 
Note that all wheels are assumed identical, hence the wheel inertia \(I_w\) and the effective tyre radius $R$ are the same between all wheels.


\subsection{Steady State Tyre Dynamics}
Tyre force models such as the non-linear Magic formula provide an accurate representation of the available road surface friction across a range of operating conditions, \emph{e.g.}, dry, wet or asphalt, cobblestone as shown in Fig.~ \ref{fig:slipmu_norm}. The non-sliding (slip ratio smaller than the optimum value) and sliding  (slip ratio larger than the optimum value) regions of the road surface are captured where the peak friction coefficient leads to the greatest tyre force and retardation of the vehicle. Thus finding the optimal wheel slip is the target of this work. Simply, the tyre force $F_{x,j}$ can be calculated using the normal load $F_{z,j}$ on the tyre:
\begin{equation}\label{eqn:tforce}
    F_{x,j} = \mu(\kappa) F_{z,j},
\end{equation}
where $\mu(\kappa)$ represents the friction coefficient for a certain road and it is a function of slip ratio $\kappa$.
According to the tyre model (\emph{i.e.}, Magic formula), it can be found by the following function \cite{Pacejka2012}: 
\begin{equation}\label{eqn:mag}
    \mu_{j}=D\sin[C\arctan({B\kappa_{j}-E(B\kappa_{j}-\arctan B\kappa_{j})})]
\end{equation}
where \(B, C, D\) and \(E\) are the stiffness factor, the shape factor, the peak factor (determining the peak friction coefficient) and the curvature factor, respectively.
In addition, the vehicle slip ratio \(\kappa_j\) is calculated as:
\begin{equation}\label{eqn:slip}
    \kappa_{j} = \frac{\omega_{j} R - U}{U}
\end{equation}
Furthermore, the practical slip ratio definition is used here giving negative force under braking as is usual practice when using the Magic formula.

\section{State and Parameter Estimation}\label{stateparamest}
Our work aims to find the maximum friction force and control the vehicle to generate such force so as to be able to stop as soon as possible.
This requires the identification of all the parameters of the magic formula (\emph{i.e.}, $\mathbf{\Theta}=[B, C, D, E]^T$), as well as the states of the ego vehicle, such as body velocity and wheel speeds. 
The \(7\)DOF vehicle dynamics represented by (\ref{eqn:fx})-(\ref{eqn:slip}) can be reduced to the nonlinear discrete-time double corner model with 3DOF, \emph{e.g.} \cite{Savaresi2010a}, as
\begin{equation} \label{eqn:sysmod}
\begin{aligned}
    \mathbf{x}_{k} & = f\left(\mathbf{x}_{k-1},{\mathbf{u}}_{k-1},\mathbf{\Theta}_{k-1} \right) \\  
    \mathbf{y}_{k} &= \mathbf{x}_{k} + \mathbf{V}_{k}
\end{aligned}
\end{equation}
where \(\mathbf{u}_k\) is the torque control input used to brake the front two wheels, \emph{i.e.}, \(\mathbf{u}_k=\left[T_{1,k},T_{2,k}\right]\), and $\mathbf{x}_k$ represents the state of the vehicle, including longitudinal body velocity and wheel speeds, \emph{i.e.}, \(\mathbf{x}_k = \left[ U_k,\omega_{f,k},\omega_{r,k} \right]\).
This is a simplified version of the previous system model whereby only the longitudinal and wheel rotational degrees of freedom are maintained. 
The wheels of each individual axle are assumed to have the same rotational velocity, hence this is a single-track model that only distinguished between front and rear slip, but not left and right.
Note that the parameter $\mathbf{\Theta}$ is assumed to be static, but may exhibit sudden step changes to match the operating condition. 
Even though the model does not include pitch and heave motions, it permits differential vertical loading conditions over the front and rear axles that can be significant during braking. 
In addition, we assume that all states are measurable but with additive noise hence matching that of the sensor, thus the system output is \(\mathbf{y}_{k} \in \mathbb{R}^{3 \times 1}\).


Note that we assume that all braking torques are applied to front wheels only as we achieve the majority of braking here. 
To estimate states and parameters at the same time, we create a single augmented vector variable $ \mathbf{\chi}_k$ as:
\begin{equation*}
    \mathbf{\chi}_k = \left[ U_k, \omega_{f,k},\omega_{r,k}, B_k, C_k, D_k, E_k \right]^T,
\end{equation*}
whose dynamics are represented by
\begin{equation}\label{eqn:abssystdyn}
\mathbf{\chi}_k = 
    \begin{bmatrix}
        \mathbf{x}_{k}\\
        \mathbf{\Theta}_{k}
    \end{bmatrix}
    =
    \begin{bmatrix}
        f\left(\mathbf{x}_{k-1},\mathbf{u}_{k-1},\mathbf{\Theta}_{k-1} \right)\\
        \mathbf{\Theta}_{k-1}
    \end{bmatrix}
    =
    \begin{bmatrix} f\left(\mathbf{\chi}_{k-1},\mathbf{u}_{k-1}\right)\\       \mathbf{\Theta}_{k-1}
    \end{bmatrix}
\end{equation}

\begin{remark}
One may take $\mathbf{\Theta}_k=\mathbf{\Theta}_{k-1}$ since we assume that all the parameters in the Magic Formula are piece-wise constants despite the changing road conditions. 
\end{remark}
\begin{remark}
Here lies the challenge associated with the tyre model identification during an ABS scenario. Our system is required to quickly determine the correct parameters whilst operating robustly to an uncertain environment. Hence the best braking action is require to be found as quickly as possible to avoid collision. 
\end{remark}

\subsection{Estimation Using Bayesian Inference}

In having a history of observations \emph{i.e.} \(\mathbf{Y}_{k-1} = \{y_0, y_1: y_{k-1}\} \) the objective of an observer is to estimate the current state \(\mathbf{\chi}_k\) using all the information available and recursively update the priori belief \(p\left(\mathbf{\chi}_k \mid \mathbf{Y}_{k-1} \right) \) as new information becomes available. 
Given we can predict the state value using our system model (\ref{eqn:abssystdyn}), one may use the Chapman-Kolmogorov equation to calculate the prior belief as:
\begin{equation}
    p\left(\mathbf{\chi}_k \mid \mathbf{Y}_{k-1}\right)=\int p\left(\mathbf{\chi}_k \mid \mathbf{\chi}_{k-1}\right) p\left(\mathbf{\chi}_{k-1} \mid \mathbf{Y}_{k-1}\right) \mathrm{d} \mathbf{\chi}_{k-1}
\end{equation}
From a new measurement at the current time step, \(y_k\), the posterior likelihood \(p\left(\mathbf{\chi}_k \mid \mathbf{Y}_{k} \right) \) can be updated using Bayes theorem based on the likelihood \(p\left( \mathbf{y}_{k} \mid \mathbf{\chi}_k \right)\), prior \(p\left( \mathbf{\chi}_k \mid \mathbf{Y}_{k-1} \right)\) and marginal likelihood \(p\left( \mathbf{y}_{k} \mid \mathbf{Y}_{k-1} \right)\):
\begin{equation} \label{eqn:prior}
    p\left(\mathbf{\chi}_k \mid \mathbf{Y}_{k} \right) = \frac{p\left( \mathbf{y}_{k} \mid \mathbf{\chi}_k \right) p\left( \mathbf{\chi}_k \mid \mathbf{Y}_{k-1} \right) }{p\left( \mathbf{y}_{k} \mid \mathbf{Y}_{k-1} \right)}
\end{equation}
where the normalising constant is:
\begin{equation}
p\left(\mathbf{y}_k \mid \mathbf{Y}_{k-1}\right)=\int p\left(\mathbf{y}_k \mid \mathbf{\chi}_k\right) p\left(\mathbf{\chi}_k \mid \mathbf{Y}_{k-1}\right) d \mathbf{\chi}_k
\end{equation}
\begin{remark}
We attempt to identify parameters of a widely accepted tyre model which, to the best of the authors knowledge, is the first complete identification effort of the Magic Formula using a Bayesian approach.
\end{remark}

\subsection{Regularized Particle Filter}
Determining the solution to (\ref{eqn:prior}) is generally intractable as it requires the entire observation history \(\mathbf{Y}_{k}\) that is costly for real online systems. In the case of linear finite dimension systems that exhibit Gaussian noise characteristics, the Kalman Filter provides the optimal recursive Bayesian solution.
However, to better approximate real systems including system non-linearities and non-Gaussian noises, a Sequential Monte Carlo (SMC) approach is more appropriate.
Due to these reasons,  the optimal state estimation is achieved by using a Bayesian solution \cite{Branko2004} that is approximated by the sum of weighted samples, \(w^i\) associated with the Dirac function \(\delta\), as the following using the number of particles, \(n_i\):
\begin{equation}
    p\left(\mathbf{\chi}_{k} \mid \mathbf{Y}_k\right) \approx \sum_{i=1}^{n_i} w_{k}^{i} \delta\left(\mathbf{\chi}_{k}-\mathbf{\chi}_{k}^{i}\right)
\end{equation}
where \(w_k^i\) is found using the importance density \(q\):
\begin{equation} \label{eqn:likeli}
    w_k^i \propto w_{k-1}^i \frac{p\left(\mathbf{y}_k \mid \mathbf{\chi}_k^i\right) p\left(\mathbf{\chi}_k^i \mid \mathbf{\chi}_{k-1}^i \right)}{q\left(\mathbf{\chi}_k^i \mid \mathbf{\chi}_{k-1}^i,\mathbf{y}_k \right)}
\end{equation}
The expected value is calculated from our particles using the standard form for finding the weighted mean:
\begin{equation}\label{eqn:exp}
    \mathbb{E} \left[\mathbf{\chi}_k\right] = \sum_{i=1}^{n_i} W_k^i \mathbf{\chi}_k^i
\end{equation}
where \(W^i_k\) is the normalised particle weight \emph{i.e.} \(W^i_k = w^i_k / \sum_{i=1}^{n_i}w^i_k\) such that the sum of the normalised particle weights equals 1.
The inference problem so far has been restricted to Sequential Importance Sampling which forms the basic mechanic for most particle filtering methods.
There are however a number of limitations to this method including degeneracy and impoverishment which must be addressed. 

The degeneracy problem exists when the weight of a single particle is updated with increasing likelihood until it reaches nearly 100\% likely where the weight, \(w_k^i\), equals 1. In this extreme scenario, computation of this single particle leads to the expected state value from the filter since all other weights are insignificant, hence the efficiency is extremely poor. 
Once degeneracy is observed in the weight values, the estimates will begin to diverge and result in unacceptably poor errors. 
To mitigate this issue, particle resampling is utilised to redistribute those with high weightings to multiple particles of the same value and remove those with low weightings. This in turn increases the number of active particles. 
Resampling methods include systematic, stratified and branch-kill with further discussions found in \cite{Li2015}.
The stratified resampling method is used in this work for its superior computational efficiency. Furthermore one may decide to selectively re-sample only when the level of degeneracy is below a threshold. The effective sample size, \(N_{eff}\), is one way to measure this \cite{Li2014} and is calculated by
\begin{equation}\label{eqn:neff}
    {N}_{eff} = \frac{1}{\sum_{i=1}^{n_i}\left(w_k^i\right)^2}
\end{equation}
Resampling is active if \(N_{eff} \leq K_o n_i \), where \(K_o\) is introduced to control the resampling frequency as in practice large value of this threshold alone leads to overly frequent resampling that can be detrimental to the estimation effort. Here we chose to define this using a linear relationship, dependant on the expected peak friction.

\begin{remark}
It is clear to see from Fig. \ref{fig:slipmu_norm} that the parameter identification problem, particularly the estimation of peak friction parameter \(D\), is more challenging on snow or icy surfaces since the peak is less defined than that for dry or wet roads. 
Therefore, one would expect a higher resampling frequency were the peak is harder to identify, ensuring an adequate distribution of particles.
\end{remark}
In this paper, $K_o$ is chosen as a time-varying parameter subject to the expectation of the road condition. 
At time step $k$, it is defined as
\begin{equation}\label{eqn:gain}
    K_{o,k} = -\frac{3}{20} \mathbb{E}\left[ D_k\right] + 0.295
\end{equation}
According to (\ref{eqn:gain}) there is no need to do manual adaptation among operating conditions and the observer is automatically adapted to the tyre model estimation of the peak friction.

Sample impoverishment is an undesirable quality of particle filtering where the particles \say{move} closer together yielding the particle population to have no diversity and is a side effect of high frequency resampling.
When all particles occupy the same value, leading to poor representation of the posterior density, this is often referred to as \say{particle collapse}. 
It can be seen that frequent resampling  would instigate this phenomenon, hence impoverishment and degeneracy are interlinked. 

To introduce particle diversity, two methods may be employed to move a particle to a new value and could be used in unison along with the resampling threshold, namely a regularization step (Regularized Particle Filter (RPF) \cite{Arulampalam2002}) and a Markov Chain Monte Carlo (MCMC) step. 
By using the RPF, jittering of the particle population is achieved.
The regularization step may be approximated to the following using the Gaussian Kernel. Note that the characteristics of the kernel are zero mean subject to small variance (\emph{i.e.}, \(\epsilon_k \sim \mathcal{N}\left(0,1\right)\)). 
The regularization step, using the re-sampled particle population, is conducted as follows: 
\begin{equation}
    \mathbf{\chi}^{i\star}_k = \mathbf{\chi}^{i}_k + h_{opt} \mathbf{D}_k \mathbf{\epsilon}^i_k
\end{equation}
where \(\mathbf{\chi}^{i\star}_k\) represents the new particle value. \(h_{opt}\) is the bandwidth calculated by 
\begin{equation}
    h_{opt} = An_i^{\frac{-1}{n_x+4}}        
\end{equation}
where $A = \left[4/ \left(n_x+2 \right) \right] ^{\frac{1}{n_x+4}}$,  \(n_x\) equals to the length of the augmented state and is selected to minimize the mean integrated square error (MISE). 
Finally, \(\mathbf{D}_k\) is the solution to the equation \(\mathbf{S}_k = \mathbf{D}_k \mathbf{D}_k^T\), where $\mathbf{S}_k$ denotes the weighted covariance and is calculated by
\begin{equation}
    \mathbf{S}_k=\frac{n_i}{n_i-1} \frac{1}{\sum_{i}^{n_{i}} W_{i,k}} \sum_{i=1}^{n_i} W_{i,k}\left(\mathbf{\chi}_{i,k}-\mathbf{\mu}_{k}\right)^{2}
\end{equation}

The purpose of the MCMC step is similar to the regularization step. However, this method ensures that a resampled particle population (or regularized sample) is free from sample degeneracy and impoverishment \cite{Li2014}. 
The aim is to draw more samples from areas with high probability by an acceptance probability, using the posterior distribution following resampling \(p(\mathbf{\chi}_k \mid \mathbf{Y}_k)\), and a proposal distribution \(p(\mathbf{y}_k \mid \mathbf{\chi}_{k}^{i\star})\).  MCMC methods randomly draw a sample (Monte Carlo) that is dependant on the likelyhood of the prior sample \emph{i.e.} Markov Chain.A common MCMC method utilised in Particle Filters is the Metropolis-Hastings Step, originally presented in \cite{Andrieu2010} with the replication in \cite{Elfring2021} utilized for this work.
\begin{remark}
The Regularization has the additional benefit of introducing artificial dynamics to the augmented state matrix. If an unlikely scenario, for example, of extreme tyre wear, very low or high road surface friction coefficients or very high vehicle speeds are observed, the particle values are able to expand to handle this.
\end{remark}

The complete regularized particle filter with MCMC step is shown in Algorithm \ref{alg:rpf}.

\begin{algorithm} 
\caption{Regularized Particle Filter with MCMC Step}\label{alg:rpf}
 \hspace*{\algorithmicindent} \textbf{Input:}  \(\mathbf{\chi}_{k-1}^i,w_{k-1}^i,\mathbf{y}_k \)\\
 \hspace*{\algorithmicindent} \textbf{Output:} \(\mathbf{\chi}_{k}^i,w_{k}^i \)
\begin{algorithmic}[1]
    \STATE Propagate system state according to  (\ref{eqn:abssystdyn})
    \STATE Determine likelihood by (\ref{eqn:likeli})
    \STATE Use  (\ref{eqn:neff}) to calculate effective ratio
    \STATE Compute the resampling gain \(K_{o,k}\) by (\ref{eqn:gain})
    \IF{\(N_{eff} \leq K_{o,k} n_i\)}
        \STATE Systematic Resampling
        \STATE Regularization
        \STATE MCMC Step
    \ENDIF
\end{algorithmic}
\end{algorithm}

\subsection{Parameter Step Change: Retrogressive Resampling}
\begin{figure}
    \centering
    \includegraphics[scale=0.9]{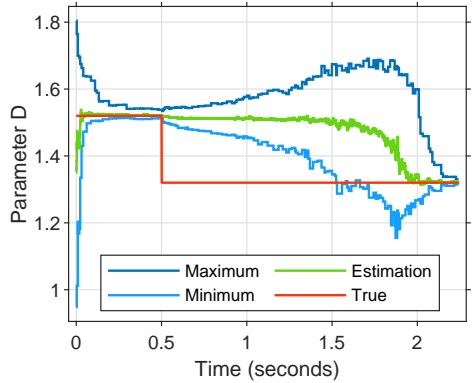}
    \caption{Estimation result on a dry road to a wet surface using the Regularized Particle Filter without retrogressive resampling. The mean of the estimation result for Parameter \(D\)  is shown with its maximum and minimum estimations.}
    \label{fig:split_D}
\end{figure}

Particle filtering is well suited to slowly changing states or parameters, such as vehicle's velocity which cannot instantaneously change from $0$ to $20m/s$ due to mechanical constraints, instead it would be accelerating over a period of time. 
However, change of road conditions is different. In fact, the changes in the road surface parameters can be best described as step changes, for example, traversing from an old road surface to a newly paved surface, from a dry road to a wet road, or over potholes. 
The performance of estimating a changing road surface is shown in Fig. \ref{fig:split_D} where an RPF estimator first converges successfully to the first road surface, but fails to quickly converge to the second one. 
It takes nearly $1.5s$ to diversify the particle population and re-converge to the new ground truth. 
The maximum and minimum particle values show the diversity and clearly indicate how the filter immediately becomes impoverished and cannot escape this effect in a timely manner. 
In fact the filter is impoverished for over a second whereby the particle population range doesn't include the true value.

To manage such sudden change in operating conditions, a simple yet effective method for resampling the particle population is introduced, namely Retrogressive resampling. 
In the case that the states of the vehicle are continuously tracked regardless of the road condition but the road surface parameters are not, a trigger to reset these parameters to their initial conditions using the current uncertainty of the system is required. 
If we observe high estimation uncertainties, a reset is triggered, leading to a normally distributed particle population over the initial range. 
The weightings are also reverted to their initial values thus giving an equal likelihood over the entire parameter space.
The Retrogresive resampling method is presented in Algorithm \ref{alg:retro} where we use 
the uncertainty of the predicted maximum tyre force, quantified by its error variance matrix, \({P}_{k+1 \mid k}\) described in Section \ref{sec:dcee}. 
This step is applied at the beginning of every new measurement.

\begin{algorithm} 
\caption{Retrogressive Resampling}\label{alg:retro}
 \hspace*{\algorithmicindent} \textbf{Input:}  \({P}_{k+1 \mid k},{\mathbf{x}}^i_k,\mathbf{\Theta}_0^i,\mathbf{\Theta}_k^i,w^i_0,P_0,w_{k}^i\)\\
 \hspace*{\algorithmicindent} \textbf{Output:} \( {\mathbf{x}}^i_k,\mathbf{\Theta}_k^i,w_k^i\)
\begin{algorithmic}[1]
    \IF{\(\max{{P}_{k+1 \mid k}} \geq P_0\)}
        \STATE \( \mathbf{\Theta}_k^i \gets \mathbf{\Theta}_0^i\)
        \STATE \({\mathbf{x}}_{k}^i \gets \mathbf{x}^i_{k}\)
        \STATE \({w}_{k}^i \gets w^i_0\)
    \ELSE
        \STATE \( \mathbf{\Theta}_k^i \gets \mathbf{\Theta}_k^i\)
        \STATE \({\mathbf{x}}_{k}^i \gets \mathbf{x}^i_{k}\)
        \STATE \({w}_{k}^i \gets w^i_k\)
    \ENDIF
\end{algorithmic}
\end{algorithm}

\begin{remark}
There is no need to tune the Retrogressive resampling process  since \(P_0\), the initial value of \(\max{{P}_{k+1 \mid k}}\), is determined by how the initial conditions are set. Following the initialisation, if at any point during the manoeuvre the uncertainty exceeds that of the initial value then we resample accordingly. 
\end{remark}

\section{Dual Control for Exploration-Exploitation}\label{sec:dcee} \label{DCEE}
The ABS problem is both a stochastic and adaptive control problem since we do not know the current state of the vehicle and parameters associated with the vehicle (tyre model) that we may exploit to achieve optimal braking.
The Particle Filter is used to find the current road surface friction \(\mu\), the maximum road surface friction \(D\), the wheel speed and body velocity. 
Thus the brake torque is selected as the action that produces the smallest tracking error between the current and the estimated maximum tyre forces.
However, this results in exploitation of the estimation solely through experience of the recursive RPF observer rather than prediction since the observer can only determine the correct parameters when it has experienced the maxima. This maxima is found by driving the vehicle to begin sliding without prediction.
It is clear that altering the brake force will change how the vehicle behaves, affecting the future state and consequently the future measurements of our vehicle state.
To this effect, the DCEE framework incorporates information from a one step prediction into our control action selection using predicted observations, propagated from the RPF observer and a discrete-time approximation of the vehicle dynamics. 
In this way, our system intelligently distinguishes those control actions that do not drive the vehicle to reach the minumum stopping distance whilst considering the uncertainty of the maximum tyre force estimation.

\subsection{DCEE Formulation}
The objective function is selected to minimize the error between the predicted tyre force \(F_{k+1 \mid k}\) and the maximum tyre force \(F_k^*\), hence is defined as:
\begin{subequations}\label{eq:2}
\begin{align}
\min _{{u}_{k}} J\left({u}_{k}\right)&=\min_{{u}_{k}} \mathbb{E}_{\mathbf{\Theta}_k}\left[\mathbb{E}_{y_{k+1|k}}\left[\big|F_{k+1 \mid k}-F_k^*\big| \mid \bar{\mathbf{\chi}}_k,\mathbf{y}_k\right]\right]\label{eq:2A}\\
\text { subject to } \notag \\
{u}_{k}&= {u}_{k-1} + {\tau}_{k} \label{eq:2B}\\
&{\tau}_{k} \in \mathcal{T} \label{eq:2c}
\end{align}
\end{subequations}
where \( \mathbf{y}_k\) and  denotes the  current measurements. \(\mathcal{T}\) is a set of control actions and \(u_k\) is the current torque which is assumed to be deterministic. 
Different from traditional control schemes, the reference is unknown and the maximum tyre force must be estimated in an online fashion according to the changes in road condition. 
Furthermore, one may expect the tyre properties to change as they are replaced due to wear during its lifespan. 
Thus an adaptive control scheme is necessary to control the vehicles braking whilst including the typical proprioceptive and exteroceptive operating dynamics of the vehicle and environment.
According to \cite{Chen2021}, the cost function (\ref{eq:2}) is equivalent to the following one using the variance of the error between the predicted tyre force and its maximum value, \({P}_{k+1 \mid k}\) :
\begin{equation}\label{eqn:dcee}
    J({u}_k) = \big| \bar{{F}}_{k+1|k} - \bar{F}_{k+1 \mid k}^*\big| + {P}_{k+1|k},
\end{equation}
where \({F}^*_{k}\) is the expectation of the predicted estimation of the total available tyre force using the peak friction coefficient \(D\), \emph{i.e.},
\begin{equation}\label{alg:expFp}
    \bar{F}^*_{k+1\mid k} = \mathbb{E}\left[F_{k+1\mid k}^{*} \mid \mathcal{Y}_{k+1 \mid k} \right].
\end{equation}
and similarly for the predicted state given a potential control action from \(\mathcal{T}\) and the estimated augmented state matrix where \(\bar{{F}}\) denotes the tyre force vector using the admissible action set \(\mathcal{T}\):
\begin{equation}\label{alg:expF}
    \bar{{F}}_{k+1 \mid k} = \mathbb{E}\left[{F}_{k+1 \mid k} \mid \mathcal{Y}_{k+1 \mid k} \right]
\end{equation}
Cost function (\ref{eqn:dcee}) shows the exploration-exploitation effect by not only incorporating the dynamic nature of the maximum tyre force but also the likelihood of achieving it, captured in the variance of the tyre force tracking error \({P}_{k+1 \mid k}\). 
Then, \(P_{k+1 \mid k}\) is found as the variance of the predicted tracking error:
\begin{equation}
    P_{k+1 \mid k} = Var \left( {\bar{{F}}}_{k+1|k} - \bar{F}_{k+1\mid k}^* \right)
\end{equation}
The first right hand term of (\ref{eqn:dcee}) exploits the current information gathered by the observer as per traditional control theory, while the second right hand term quantifies the uncertainty of the vehicle reaching the predicted maximum tyre force.

\begin{remark} 
ABS is fundamentally an exploration-exploitation problem since the maximum friction of a road surface is unknown and cannot be measured. Thus the control system should aim to understand the environment by means of the available actions (\emph{exploration}) observing the response, then taking the best action that results in the shortest stopping distance (\emph{exploitation}) in a recursive manner. Most existing methodologies do not acknowledge this problem instead, computing the most appropriate action from the posterior belief of the road (exploitation only). SMC, for example, only considers the reference surface rather than exploring the system to find it.
\end{remark}

\subsection{Measurement Prediction}
The goal of the measurement prediction is to determine the actions that achieve the highest brake force, before taking them. Since we have already generated a belief in the current state of the vehicle one may predict what the next tyre force measurement is, given a selected control input. 
In practice, physical measurement of the tyre force represents a significant challenge since unconventional sensors, such as Load Sensor Bearings \cite{Nam2013} or other Smart Tyre systems \cite{Mazzilli2021}, are used. 
However this work does not require such systems to estimate the tyre force or predict it.

\begin{remark}
    This work does not require sophisticated sensing, instead estimating the current and predicted tyre force from low cost conventional sensors.
\end{remark}

The predicted tyre force for $i^{th}$ particle, \({{F}}_{k+1 \mid k}^i\), is found using a discrete time approximation of the half car dynamics using the posterior estimate of the augmented state matrix, \(\mathbf{\chi}_{k}^i\) from the RPF and the vertical load on the wheel:
 \begin{equation}
    {{F}}_{k+1 \mid k}^i = \mu_{k+1|k}^i F_{z}
\end{equation}
where \(\mu_{k+1 \mid k}^i\) is the predicted friction coefficient found using the posterior estimation of the augmented state \emph{i.e.} \(\bar{\mathbf{\chi}}_{k}\). Then similarly for the predicted maximum tyre force using peak factor \(D\) where \(D_{k+1\mid k}^i= D_k^i\) instead of the predicted friction coefficient.

This however leads to \(n_i\) predicted tyre forces for each action 
which is unnecessarily large since the performance gain of \(n_i\) particles (potentially thousands of measurements for each action) does not justify the computational cost. Therefore one may `intelligently' select the most relevant observations using the stratified resampling technique \cite{Li2015} with the posterior weighting. 
This facilitates the redistribution of the particle population into strata or within sub-populations that are drawn randomly from within these groups. Considering that the prediction is inherently uncertain, we determine that only \(40\) predicted observations is required for ABS. 
Performing resampling on the predicted particle population also has the additional benefit of ensuring that the observations are non-degenerate thus ensuring a set of unique measurements.
Finally, making the assumption that the tyre force observations follow a Gaussian distribution with variance \(\sigma_f=250\), one determines the likelihood of the predicted tyre force. Ensuring that when approximating the tyre forces by Eqs.~(\ref{alg:expFp})-(\ref{alg:expF}), we use the down sampled set of predicted tyre force measurements when determining the weights.

\section{Simulation} \label{simulation}
The effectiveness of the aforementioned closed loop control scheme is demonstrated in simulation using the system architecture shown in Fig.  \ref{fig:dceearch} with the 7DOF vehicle model (\emph{i.e.}, (\ref{eqn:fx})-(\ref{eqn:slip})) parameterised using a Jaguar XJ vehicle as shown in Table \ref{table:1}. 
\begin{table}[h]
\caption{Vehicle parameters.}
\begin{center}
\begin{tabular}{|m{4.1cm}| m{1cm}| m{1cm} |}
    \hline
    Parameter & Value & Units\\
    \hline
    Front Suspension Stiffness, \(k_f\) & 20090 & \(N/m\) \\
    \hline
    Rear Suspension Stiffness, \(k_r\) & 22700 & \(N/m\) \\
    \hline
    Front Damping Coefficient, \(c_f\) & 2000 & \(N/m^2\) \\
    \hline
    Rear Damping Coefficient, \(c_r\) & 2260 & \(N/m^2\) \\    
    \hline
    Rolling Radius, \(R\) & 0.29 & \(m\) \\
    \hline
    Vehicle Body Inertia, \(I_{yy}\) & 3983 & \(kg/m^3\) \\
    \hline
    Center of gravity, \(x_G\) & 0 & \(m\)\\
    \hline
    Center of gravity, \(y_G\) & 0 & \(m\)\\
    \hline
    Center of gravity, \(z_G\) & -0.4427 & \(m\)\\
    \hline
    Distance to front axle, \(a\) & 1.455 & \(m\) \\
    \hline
    Distance to rear axle, \(b\) & -1.575 & \(m\) \\
    \hline
    Vehicle Total Mass, \(m_T\) & 1838.35 & \(kg\) \\
    \hline
    Wheel Inertia, \(I_w\) & 1.25 & \(kg/m^3\) \\
    \hline
\end{tabular}
\end{center}
\label{table:1}
\end{table}
Online  estimation of the augmented state vector $\mathbf{\chi}_{k}$ is done in all scenarios where the road surface friction is modelled using the Magic formula  with the parameters of the road surface shown in Table \ref{table:2}. It is acknowledged that the tyre stiffness changes between road types demonstrating the insensitivity of this method to different tyres thus further showing the advantages of the DCEE based solution. Measurements of the ego body velocity and wheel speed {$[U,\omega_f,\omega_r]^T$} are corrupted by the additive noise $\mathbf{V}_k$ such that \( \mathbf{V}_k \sim \mathcal{N}\left(\left[0,0,0\right]^T, diag\left(0.2,0.5,0.5\right)\right)\).
\begin{table}[h]
\caption{Magic formula coefficients and optimal slip ratio for three road surfaces.}
\begin{center}
\begin{tabular}{ | m{5.2em}| m{1.5cm}| m{1.5cm}| m{1.5cm}|} 
  \hline
  Parameter & Dry Road & Wet Road & Snowy Road \\ 
  \hline
  \(B\) & 5 & 10.695 & 20\\ 
  \hline
  \(C\) & 1.4601 & 1.4 & 1.5354\\ 
  \hline
  \(D\) & 1.3 & 0.8 & 0.3\\ 
  \hline
  \(E\) & -10.3522 & -3.5 & 0.8525\\ 
  \hline
  Optimal Slip & -0.15 & -0.104 & -0.18\\
  \hline
\end{tabular}
\label{table:2}
\end{center}
\end{table}

In real world scenarios and in the context of ABS, the ego vehicle begins in an unknown environment, particularly if there has been rain or snow for example. 
Therefore the particles are initialised with a set of parameter values that encompass all road types and scenarios mentioned here.
These are represented as uniform distributions: \(B \in \mathcal{U}\left[4,21\right]\), \(C \in \mathcal{U}\left[1.2601,1.601\right]\), \(D \in \mathcal{U}\left[0.2,1.6\right]\) and \(E \in \mathcal{U}\left[-12,2\right]\). 
Additionally, since we assume the vehicle to be in an unknown environment we also assume proprioceptive uncertainty where we find the initial states to found from any low cost sensor then uniformly distributed as per the quality of the sensor. Although some sensors exhibit Gaussian noise, we do not need to make such assumption given our observer design. 
In the case of body velocity and wheel speed, they are within [\(v_0 -2m/s\), \(v_0+2m/s\)] of the first GPS velocity measurement and [\(\omega_0-3.5 rad/s\), \(\omega_0+3.5 rad/s\)] from the wheel speed sensor.

\subsection{Change In Initial Speed}
\begin{figure}
    \centering
    \includegraphics[scale=0.85]{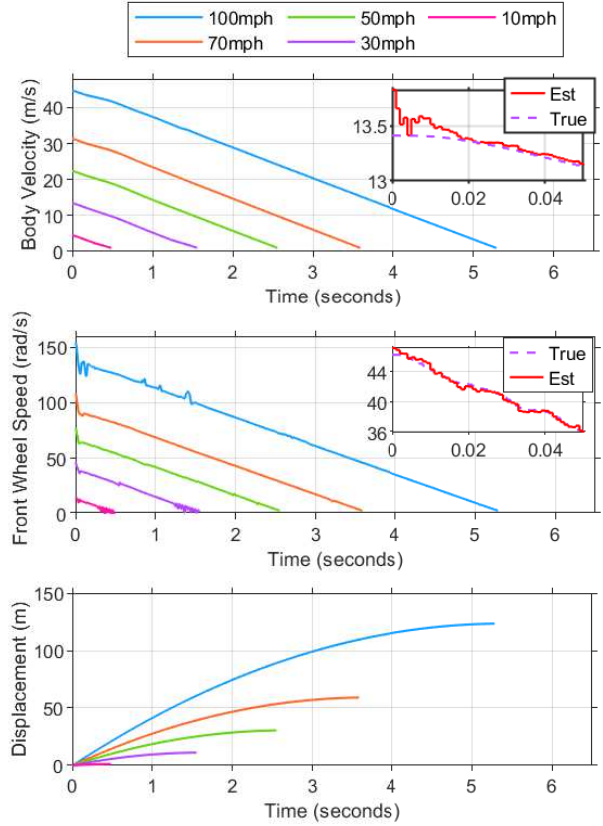}
    \caption{Simulation results with different starting velocity, braking to stationary. The body velocity and wheel speed estimation result is also shown for the 30mph braking scenario.}
    \label{fig:speed}
\end{figure}

In reality, the ABS should work in a wide range, from high to low speeds, including those above the normal driving speeds \emph{i.e.} above the typical speed limit.
Therefore, we tested the proposed DCEE based solution for a range of initial velocities, from $10 mph$ to $100 mph$, and aimed to stop the vehicle as soon as possible. Simulation results are  shown in Fig.  \ref{fig:speed}. Beneficial to this technique is that no gain tuning is required between operating points \emph{i.e.}, the same model, as well as the same parameter values, are used for each starting velocity. No locking of the wheel occurs and the vehicle uniformly decelerates over all velocities. 
One notices the oscillation in the wheel speed at low body velocities when staring from $30$ and $10 mph$. This may be attributed to the low number of steps the controller has taken where the PF does not converge as well therefore not holding as steady wheel deceleration. Although this is only for a very short period of $0.2 s$. 
Furthermore, initial variation, particularly in the \(100mph\) test case, is a result of the the controller uncertainty in the location of the maximum tyre force. This period is short since the controller can quickly identify and reach the appropriate tyre force. 
Further oscillations in the wheel speed are due to the observer resampling which is always greater in the initial period as the filter converges to the true state/parameter, but exist throughout the manoeuvre. This is required to prevent degradation in the filter and improve the identification of less sensitive parameters.

Very quick convergence of the ego states is achieved as is shown on Figure \ref{fig:speed}. In fact less than 0.5\% error is observed in all vehicle states after 13 time steps however the tyre model parameters requires a longer time and is aided throughout the scenario by the excitation caused through the explorative effort of the DCEE controller. 

\subsection{Adverse Road Conditions}
Road conditions vary a lot, from dry to muddy and from wet to snow. 
Furthermore, the road surface does not always remain the same. For example, water may collect on parts of the road but drain away on other sections of the road, thus ABS should be able to work efficiently and adapt to these changes. 
The robustness of our ABS solution to both static and changing road conditions is tested in this sub-section.


\subsubsection{Static Road Surface Estimation}
\begin{figure}
    \centering
    \includegraphics[scale=0.75]{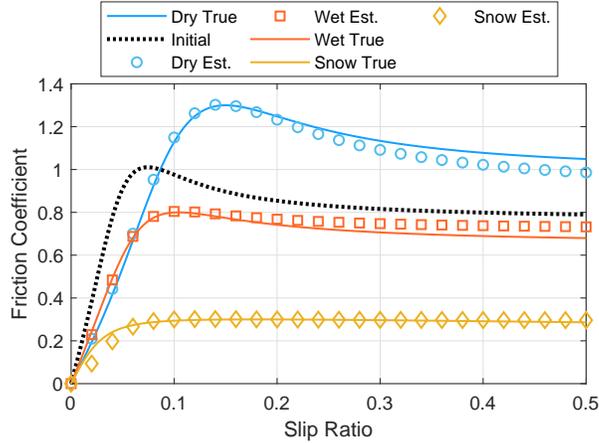}
    \caption{Parameter estimation result during three separate ABS braking manoeuvres with time-invariant road conditions and same initial conditions.}
    \label{fig:roadcond}
\end{figure}

The DCEE method is well suited to the online identification of many road surfaces with the estimation result of dry, wet and snow covered road surfaces during an ABS manoeuvre from \(20m/s\) shown in Fig.~\ref{fig:roadcond}. The road surfaces are identified correctly, estimating the initial non-sliding elastic region and defining the maxima appropriately.  Small errors found at higher slips where the vehicle has had limited experience for, \emph{i.e.}, driving the vehicle to a condition when the slip ratio equals to 0.4 would cause significant sliding thus the vehicle never reaches this.  Therefore due to the lower exploration effort, the identification performance becomes degraded.
However, as the goal of the ABS is to keep the vehicle working at the condition when the friction coefficient achieves the maximum value, the results are more than acceptable.

\subsubsection{Dynamic Road Surface}
\begin{figure}
    \centering
    \includegraphics[scale=0.9]{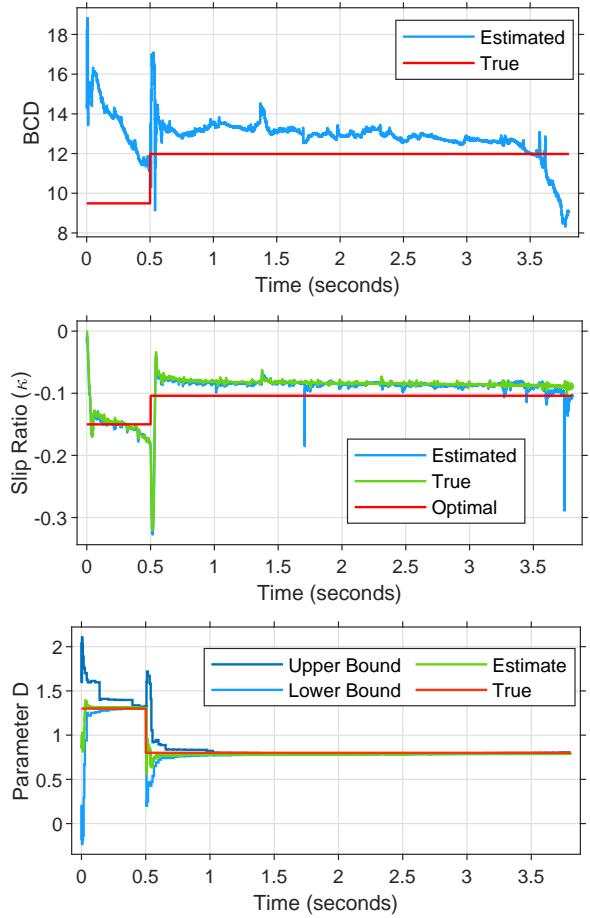}
    \caption{Estimation result of tyre model parameters.}
    \label{fig:tyrest}
\end{figure}

The RPF estimation results of individual parameters for a time-varying road surface (the road surface changes from dry to wet at $0.5 s$) can be found in Fig.  \ref{fig:tyrest} where the sensitivity of the parameters is made clear. 
Parameter \(D\) converges almost immediately and tracks the step change very well. Comparing this to the original result (that does not use Retrogressive resampling) in Fig.~\ref{fig:split_D}, one can clearly conclude the importance of the Retrogressive resampling approach. 
The remaining parameters have mixed tracking performance with parameter E showing to be insensitive to changes throughout the surface but without any detrimental impact on the controller or vehicle since this parameter governs the sliding region of the tyre model which the ABS prevents the vehicle from operating in. 
More importantly, one may look to the product of the parameters \(B\), \(C\) and \(D\) since this defines the gradient of the slip-\(\mu\) curve and by taking the arcsine of this product one finds the angle of the slope. Shown in Fig.~\ref{fig:tyrest}, one can observe that this estimation is slower than that for parameter \(D\), eventually converging by the end of the manoeuvre.

\begin{figure}
    \centering
    \includegraphics[scale=0.78]{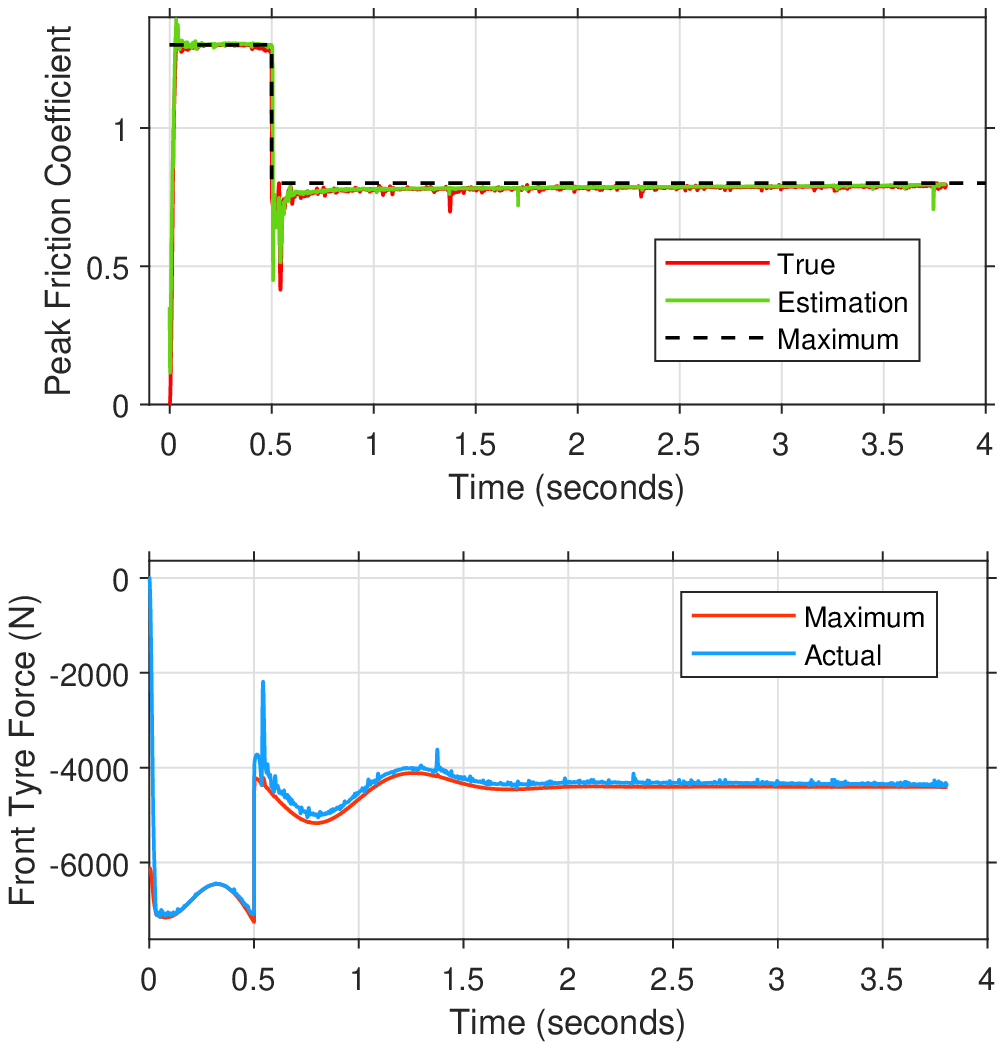}
    \caption{Brake manoeuvre when road surface changes from dry to wet road at $0.5$ seconds.}
    \label{fig:splitmu}
\end{figure}

The vehicle behaviour during the braking procedure can be assessed by tyre forces, as shown in Fig.~\ref{fig:splitmu}. The maximum available tyre force is time-varying. Thus, the weight transfer during the manoeuvre is also time-varying.
However, the DCEE method is able to continually drive the vehicle to this varying operating condition  without lock of any wheels. Estimation of the current time-varying friction coefficient at the wheels closely tracks the true friction (see Fig.  \ref{fig:splitmu}) along with holding a steady state error of less than 0.02 (\emph{i.e.}, $\leq3\%$) with maximum available.

\begin{figure}
    \centering
    \includegraphics[scale=0.77]{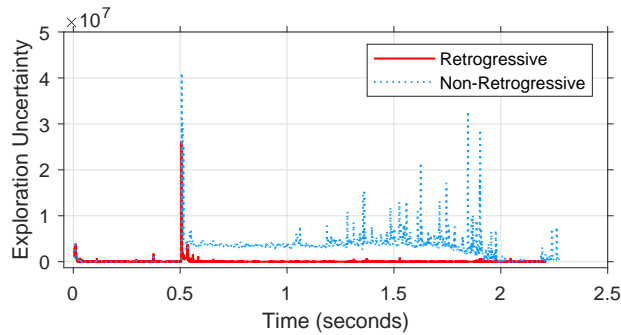}
    \caption{Effect of retrogressive resampling on control system uncertainty.}
    \label{fig:retro}
\end{figure}

It can be shown that by including the Retrogressive resampling step one not only improves the augmented state estimation but also reduces the uncertainty of the location of the maximum available tyre force. 
Shown in Fig.~\ref{fig:retro} is a comparison between the same changing road surface scenario with and without the resampling step. 
By using this method, one reduces the estimation uncertainty (especially after $0.5 s$ when the vehicle is driven on a wet road) and provides a quicker (\emph{i.e.}, within $0.1 s$) convergence to the steady state.

\subsection{Comparison with Extremum Seeking Control}
\begin{figure}
    \centering
    \includegraphics[scale=0.85]{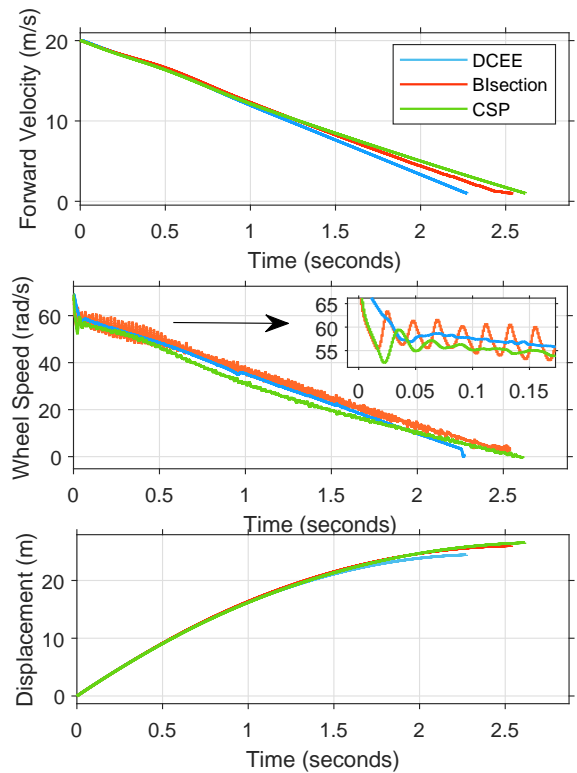}
    \caption{Comparison between the DCEE method and two Extremum seeking methods, stopping from 20m/s (45mph).}
    \label{fig:compari}
\end{figure}

\begin{figure}
    \centering
    \includegraphics[scale=0.75]{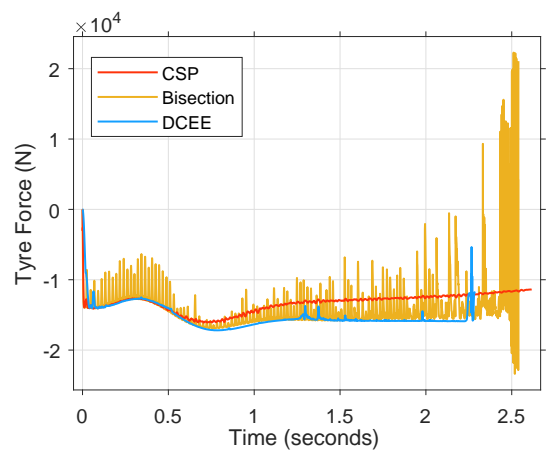}
    \caption{Total vehicle tyre force comparison during braking manoeuvre from \(20m/s\).}
    \label{fig:maxcomp}
\end{figure}

To demonstrate the superiority of our DCEE method above existing schemes, we make comparison with two Extremum Seeking Control (ESC) for the ABS problem, referred to as Continuous Sine Perturbation (CSP) \cite{Ariyur2003} and Bisection method (Bisection) \cite{Xu2020}. Having identified the optimal slip ratio we utilise the P-controller and reachability based control schemes also presented in these works. The P-controller uses the gain \(c=500\), chosen to match the performance for our vehicle whereas all other parameters for the CSP model remain the same. The sampling time is, \(\Delta T = 0.001\) and the initial conditions for the Bisection method are: \(a=0.1\), \(b=0.2\) and tolerance is \(l_d = 0.001\). 
Measurements of the ego velocity and wheel speed are made in all systems whereby a low pass filter with cutoff frequencies of \(20Hz\) and \(100Hz\) for ego body and wheel speed sensors respectively are used to remove the noise except in the DCEE approach where the aforementioned Particle Filter is used. Simulation results are displayed in Figs.~\ref{fig:compari} and \ref{fig:maxcomp}. 

In this example, the DCEE approach is compared  to both ESC methods, stopping the vehicle in $2.274 s$ which is $15.0\%$ faster than the CSP method and $11.7\%$ faster than the bisection method. This leads to an overall stopping distance of $24.45 m$ that is $2.09 m$ shorter than that of CSP and $1.56 m$ than that of the bisection method. Noticeably, the reachability controller used in the bisection method is unable to maintain a stable wheel speed and tends away from constant vehicle deceleration. 
Furthermore, one may notice slightly slower deceleration of the wheel speed which can be attributed to the DCEE being primarily driven by the exploration of the state and parameter space. This however is beneficial since the wheel speed in our method begins a steady deceleration much faster than either of the ESC methods and with no chatter.
Similarly the other methods also go through the explorative phase although are only identifying the optimal slip, not the road condition as and optimal tyre force as we are. 
The bisection method has large oscillations, undershooting the correct wheel speed and the CSP method undershoots but at steady state, reaching a slower deceleration thus causing higher wheel speeds.

Noticeably the vehicle with DCEE is able to achieve a greater total tyre force given the transient loading condition and is highlighted in Fig.  \ref{fig:maxcomp}. 
The inclusion of the suspension dynamics is imperative for simulation and shows the limitation for classical adaptive control, such as the CSP, fails to maintain as large tyre force throughout the transient loading behaviour of the vehicle. 
Conversely the Bisection method is able to track this change although with significant instability, extending into the steady state part of the manoeuvre. 
Our DCEE method achieves and maintains a greater total tyre force with superior stability even while identifying the tyre model parameters which the other methods do not.

\section{Conclusion} \label{conclusion}
The DCEE method has been demonstrated as an effective ABS solution. Estimation of vehicle states from noisy sensors and identification of Magic formula tyre model parameters are done so within this framework in an online manner. We have extended the use of the particle filtering method to parameter identification in the presence of sudden changes in operation using the Retrogressive resampling method which incorporates the estimation uncertainty of the maximum available tyre force in a holistic, goal-orientated way. 
This work shows a performance margin of up to $15\%$ in stopping time and $8.5\%$ in stopping distance compared to existing ABS methods while achieving a higher steady state braking force.

Future work may include introducing a longer prediction horizon to the DCEE methodology to bring greater stability in the control action although would undoubtedly introduce higher computational cost. Whilst this method considers unknown vehicle states and unknown vehicle parameters, extending this framework to identify the transient tyre behaviour would also be beneficial.

\bibliographystyle{IEEEtran} 
\bibliography{references}

\end{document}